\documentstyle[prl,aps,multicol,epsf]{revtex}
\begin{document}
\draft
\widetext
\title{Paramagnetic-ferromagnetic transition in a double-exchange model}
\author{Eugene Kogan$^{1}$ and Mark Auslender$^2$}
\address{$^1$ Jack and Pearl Resnick Institute 
of Advanced Technology,
Department of Physics, Bar-Ilan University, Ramat-Gan 52900, 
Israel\\
$^2$ Department of Electrical and Computer Engineering,
Ben-Gurion University of the Negev,
P.O.B. 653, Beer-Sheva, 84105 Israel}
\date{\today}
\maketitle
\begin{abstract}
\leftskip 54.8pt
\rightskip 54.8pt
We study paramagnetic - ferromagnetic transition  due to
exchange interaction between  classical localized magnetic moments
and conduction
electrons. We formulate the Dynamical Mean
Field Approximation equations for arbitrary 
electron dispersion law, concentration  and  relation
between  exchange coupling and the electron band width. Solving these
equations we  find  explicit formula for the transition temperature $T_c$. 
We present the
results of calculations of the $T_c$ 
for the semi-circular electron density of states.
 
\end{abstract}
\begin{multicols}{2}
\narrowtext

\section{Introduction}

The  double-exchange (DE) 
model \cite{zener,anderson,degennes} is one of the basic ones 
in the theory of magnetism. Magnetic ordering appears in this model  due to   
exchange coupling  between the core spins and the 
conduction electrons. The Hamiltonian  of the model is 
\begin{eqnarray}
\label{HamDXM}  
H = \sum_{nn'\alpha} t_{n-n'} c_{n\alpha}^{\dagger} c_{n'\alpha}
-J \sum_{n\alpha\beta} {\bf S}_n\cdot 
{\bf \sigma}_{\alpha\beta}c_{n\alpha}^{\dagger} c_{n\beta},
\end{eqnarray}
where $c$ and $c^{\dagger}$ are the electrons annihilation and creation
operators, ${\bf S}_n$ is the operator of a core spin,  $t_{n-n'}$ is the 
electron hopping,  
$J$ is  the 
exchange 
coupling between a core spin and n electrons,
$\hat{\bf \sigma}$ is the vector of the Pauli matrices, and $\alpha,\beta$ are
spin indices.

The model (the core spins being treated as classical
vectors) began to be studied  several decades ago
 \cite{zener,anderson,degennes}.  Because of a 
recent general
interest in manganites, it reappeared in the
focus of attention (see reviews \cite{coey99,izyumov01,ziese02,edwards} 
and references therein).  In particular, 
application  of the Dynamical Mean 
Field Approximation (DMFA) \cite{DMFA} substantially advanced our
understanding of the properties of the DE model
\cite{furukawa,furukawa2}. 
Most of
the papers dealing with the  model, starting from classical paper by De
Gennes \cite{degennes}, considered the DE Hamiltonian
with infinite exchange (and  with the addition of the antiferromagnetic
superexchange, which is crucial for the explanation of magnetic properties of
manganites). However lately many properties of the model for arbitrary strength
of the exchange were analyzed \cite{chat}.

In this paper we calculate the temperature  of a paramagnetic-ferromagnetic  
transition $T_c$ in a
double-exchange model  for arbitrary 
electron dispersion law, concentration  and  relation
between the exchange coupling and the electron band width by formulating and 
solving
the DMFA equations. 
We treat the
core spins as classical
vectors.

\section{DMFA equations and $T_c$}

Spins being considered as classical
vectors $ {\bf S}_n = {\bf m}_n$ 
(with the normalization $|{\bf m}|^2 = 1$), the problem becomes a single
electron one. The DE Hamiltonian 
 in a single electron representation 
 can be presented as
\begin{equation}
\label{generic}
H_{nn'}=t_{n-n'}-J {\bf m}_n\cdot {\bf \sigma}\delta_{nn'}.
\end{equation}
Let us introduce Green's function
\begin{eqnarray}
\label{green}
\hat{G}(E)=(E-H)^{-1} 
\end{eqnarray}
and local Green's function 
\begin{eqnarray}
\hat{G}_{\rm loc}(E)= 
\left\langle\hat{G}_{nn}(E)\right\rangle.
\end{eqnarray}
In the last Equation, the averaging is with respect to random configurations of
the core spins. 
In the framework of the DMFA approach to the problem (see \cite{DMFA,furukawa2} 
and references therein) the local Green's function
is expressed through the  the local self-energy $\hat{\Sigma}$ by the
equation
\begin{eqnarray}
\label{local}
\hat{G}_{\rm loc}(E) =g_0\left(E - \hat{\Sigma}(E)\right),
\end{eqnarray}
where
\begin{eqnarray}
\label{g}
g_0(E) =\frac{1}{N}\sum_{\bf k}\left(E-t_{\bf k}\right)^{-1} 
\end{eqnarray}
is the bare (in the
absence of the  exchange interaction) local Green's function. The
self-energy satisfies equation
\begin{eqnarray}
\hat{G}_{\rm loc}(E)=\left\langle \frac{1}
{\hat{G}_{\rm loc}^{-1}(E)+\hat{\Sigma} (E)
+J{\bf m}\cdot\hat{\bf \sigma}}\right\rangle,
\label{cpa}
\end{eqnarray}
where $\left\langle X({\bf m})\right\rangle \equiv \int X({\bf m})P({\bf m})$,
and $P({\bf m})$ is a probability 
of a given spin orientation (one-site probability). The quantities 
$\hat{G}$ and $\hat{\Sigma}$
are $2\times 2$ matrices in spin space.
 
The difference between Eq. (\ref{cpa}) and similar equation 
appearing in the Coherent Potential Approximation \cite{ziman},
is due to the fact that in our problem the disorder is annealed (and not
quenched). That is the probability of a given
(random) core spin configuration is determined by the energy of electron
subsystem interacting with this configuration. 

To understand the DMFA assumption
allowing to find the probability $P({\bf m})$ 
self-consistently with the solution of Eq. (\ref{cpa}), notice that
Eq. (\ref{cpa})  
reduces the problem of electron scattering by many spins,
to  scattering by a single spin  with  the effective 
potential 
$V=-J {\bf m}\cdot {\bf \sigma}-\hat{\Sigma}$. The spin is 
embedded in an effective medium, 
described by the Hamiltonian $t_{\bf k}+\hat{\Sigma}$,
and, hence, by the local Green's function $\hat{G}_{\rm loc}$. 
Consider the change in the number of electron states in such system
due to a single spin. We get   \cite{ziman,hewson} 
\begin{eqnarray}
\label{probability4}
\Delta D(E,{\bf m})= -\frac{1}{\pi}{\rm Im}
\ln {\rm det}\left[1+\left(J{\bf m}\hat{\bf \sigma}
+\hat{\Sigma}\right)\hat{G}_{\rm loc}\right],
\end{eqnarray}
where the argument of both $G_{\rm loc}$ and $\Sigma$ is $E+i0$. 
So the change in thermodynamic potential is  \cite{doniach,chat,ak2}
\begin{eqnarray}
\Delta\Omega({\bf m})=\int f(E)\Delta D(E,{\bf m})dE,
\label{probability}
\end{eqnarray}
where  $f(E)$ is the Fermi function.
The  
DMFA approximation  for the one-site
probability
$P({\bf m})$ is:
\begin{eqnarray}
\label{prob2}
P({\bf m})\propto \exp\left[-\beta\Delta\Omega({\bf m})\right].
\end{eqnarray}

Eqs. (\ref{cpa}) and (\ref{prob2}) are the system of  non-linear (integral)
equations. However, the system is grossly simplified 
in paramagnetic (PM) phase, where the macroscopic magnetization
${\bf M}=0$. In this case $P({\bf m})={\rm const}$,  
the averaging in Eq. (\ref{cpa}) can be 
performed explicitly, and we obtain
\begin{eqnarray}
g(E)= \frac{1}{2}\sum_{(\pm)} \frac{1}
{g^{-1}(E)+\Sigma(E)
\pm J},
\label{cpa2}
\end{eqnarray}
where   
$\hat{\Sigma}=\Sigma\hat 1$, $\hat{G}=g\hat 1=g_0(E-\Sigma)\hat 1$, 
where $\hat 1$ is a unity matrix.

In the ferromagnetic (FM) phase near the
Curie temperature, Eqs. (\ref{cpa}) and (\ref{prob2}) can be linearized
\cite{furukawa,chat} with respect to 
${\bf M}$. Thus 
we  reduce the DMFA equations
to a traditional MF equation  \cite{aus02}
\begin{eqnarray}
P({\bf m})\propto  \exp\left( -3\beta T_c{\bf M}\cdot{\bf m}\right).
\label{probability2}
\end{eqnarray}
The parameter $T_c$ is formally introduced as a coefficient in the 
expansion of $\Delta \Omega({\bf m})$ with respect to
${\bf M}$.
Non-trivial solution of the MF equation 
\begin{equation}
{\bf M}=\left\langle{\bf m}\right\rangle
\end{equation}
can exist only for $T<T_{c}$, hence $T_c$ is the ferromagnetic transition
temperature. 
In all cases $T_c$ turns out to be much less the chemical potential, so we
can consider electron gas as degenerate. In this case
\begin{equation}
\int_{-\infty}^{\infty}f(E)\dots dE\equiv\int_{-\infty}^{\mu}\dots dE,
\end{equation}
where the chemical potential (in paramagnetic phase) is found from the equation
\begin{equation}
n=-\frac{2}{\pi}\int_{-\infty}^{\mu}{\rm Im}\;g \; dE,
\end{equation}
and $n$ is the number of electrons per site.

For the $T_c$, after straightforward, though lengthy algebra, we
obtain
\begin{eqnarray}
\label{Theta}
T_{c}=\frac{2J^2}{3\pi}\int_{-\infty}^{\mu}
\mbox{Im}\left[\frac{g}{
\frac{(\Sigma g'-g)(1+\Sigma g)}
{g'+g^2} -\frac{2J^2g}{3}}\right]dE,	
\end{eqnarray}
where $\Sigma$ and $g$ are determined by the properties of the system in
the PM
phase; they are the solutions of Eqs. (\ref{local}) and
(\ref{cpa2}); 
$g'\equiv\left.\frac{d g_0(\varepsilon)}
{d \varepsilon}\right|_{\varepsilon=E-\Sigma}$.
Eq. (\ref{Theta}) is  the main result of our paper.

In the case of weak exchange $J\to 0$
Eq. (\ref{Theta}) takes the form
\begin{eqnarray}
\label{rkky2}
T_{c} = \frac{2J^2}{3}\left[N_0(\mu)-\frac{1}{\pi }\int_{-\infty}^{\mu}
\mbox{Im}\;g_0^2(E)\right]dE.
\end{eqnarray}
In fact, this equation  is the MF approximation 
\cite{aus02} for the RKKY Hamiltonian \cite{vf}, to which the original
Hamiltonian
(\ref{HamDXM}) can be reduced  in the case $J\ll W$. 

In the opposite case $J\to \infty$ we have to consider only one spin sub-band.
(Because of the symmetry of the problem with respect to transformation
$n\rightarrow 2-n$, everywhere further on we consider  only the case $n\leq 1$.)
After we shift $E$, $\Sigma$ and $\mu$ by $J$, we obtain
\begin{eqnarray}
\label{Theta2}
T_{c}=\frac{2}{\pi}\int_{-\infty}^{\mu}
\mbox{Im}\left[\frac{g'+g^2}{g'-2g^2}\right]dE,	
\end{eqnarray}
where $g$ is found from the equation
\begin{eqnarray}
g(E)= \frac{1}{2}\ \frac{1}
{g^{-1}(E)+\Sigma(E)}.
\label{cpa3}
\end{eqnarray}

\section{$T_c$ for semi-circular DOS}

Let us apply Eq. (\ref{Theta}) to the case of  
semi-circular (SC) bare density of states (DOS)
\begin{equation}
N_0(\varepsilon)=\frac{2}{W}\sqrt{\left(\frac{E}{W}\right)^{2}-1}.
\end{equation}
Then
\begin{eqnarray}
\label{gint}
g_0(E)=\int \frac{N_0(\varepsilon)d \varepsilon }{E - \varepsilon}
=\frac{2}{W}\left[\frac{E}{W}-
\sqrt{\left(\frac{E}{W}\right)^{2}-1}\right].
\end{eqnarray}
From Eq. (\ref{local}) follows
\begin{equation}
\label{sigma}
\Sigma=E-W^2g/4-g^{-1},
\end{equation}
and Eq. (\ref{cpa2}) can be written as
\begin{eqnarray}
\label{rq2}
g=\frac{1}{2}\sum_{(\pm)}\frac{1}{E-W^2g/4\pm J}.
\end{eqnarray}
Eq. (\ref{Theta}) takes the form
\begin{eqnarray}
\label{Thetasc}
T_{c}=\frac{J^2W^2}{6\pi}\int_{-\infty}^{\mu}\nonumber\\
\mbox{Im}\left[\frac{g^2}
{\left(E-\frac{W^2g}{4}\right)\left(E-\frac{W^2g}{2}\right) -
\frac{J^2W^2}{6}g^2}\right]dE.	
\end{eqnarray}
In the case of weak exchange $J/W\ll n^{1/3}$, we obtain
\begin{eqnarray}
\label{rkky3}
T_{c} = \frac{4J^2}{3\pi^2 W}(4y^2-1)\sqrt{1-y^2},
\end{eqnarray}
where
\begin{equation}
n=n_0(y)\equiv 1+\frac{2}{\pi}\left(\sin^{-1}y+y
\sqrt{1-y^2}\right).
\end{equation}

In the  case of strong exchange $J/W\gg n^{1/3}$, Eq. (\ref{Theta2}) takes the form \cite{furukawa2,ak2}
\begin{eqnarray}
\label{theta}
T_{c}=\frac{W^2}{6\pi}\int_{-\infty}^{\mu}
\mbox{Im}\left[\frac{ g^2}{1 - \frac{W^2g^2}{6}}\right]dE\nonumber\\	
=\frac{W\sqrt{2}}{4\pi}\left[ \sqrt{1-y^2}-\frac{1}{\sqrt{3}}\tan^{-1} 
\sqrt{3(1-y^2)}\right],
\end{eqnarray}
where $n=n_0(y)/2$. If, in addition, $n\ll 1$, we obtain
\begin{equation}
T_c=\frac{3}{4\sqrt{2}}Wn\approx .53\; Wx.
\end{equation}
Notice, that virtual crystal approximation (for $J\to\infty$) \cite{degennes}  
gives 
\begin{equation}
T_c=\frac{2}{15}Wx\approx .13\; Wx,
\end{equation}
thus being imprecise by a factor of 4.

For arbitrary exchange, the integral in Eq.
(\ref{Thetasc}) can be calculated only numerically.  The results of such
calculation are presented on FIG. 1.
\begin{figure}
\epsfxsize=3truein
\centerline{\epsffile{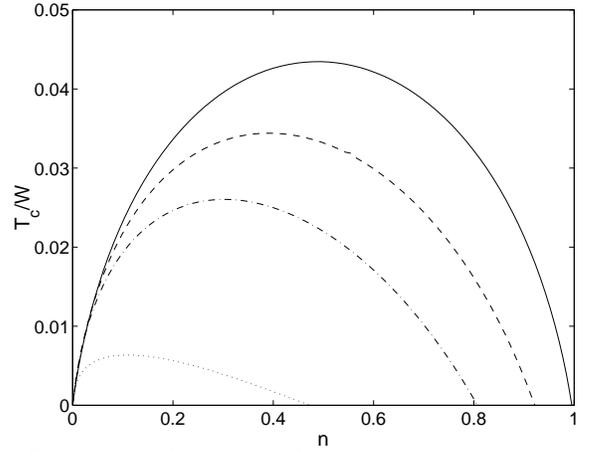}}
\label{FIG.1}
\caption{$T_c$ as a function of
 electron concentration $n$ for different relative strengths of the exchange:
  $J/W=.25$ (dotted line), $J/W=1$ (dash-dotted
 line),  $J/W=2$ (dashed line), and $J/W=20$ (solid line).} 
\end{figure}

\section{Nonferromagnetism}

Note that  that our main
result (equation for the $T_c$) indicates it's own 
limits of validity. 
In part of the $J/W-n$ plane,  Eq. (\ref{Thetasc})
gives $T_c<0$.
Negative value of $T_c$
means, that  at
any temperature, including
$T=0$,  the paramagnetic phase 
is  stable   
with respect to  appearance of small 
spontaneous magnetic moment,  and
strongly suggests that the ground state in
this part of the phase plane is nonferromagnetic 
\cite{chat3}. The mechanisms leading to  
the destruction of ferromagnetism can be easily understood in the extreme cases
of weak and strong exchange.

It is well known, that for  $J\to 0$ the DE Hamiltonian  can be reduced to
to the RKKY Hamiltonian (see review \cite{vf} and
references to the original papers therein)
\begin{eqnarray}
H_{RKKY}=\sum_{nn'}I_{nn'}{\bf
m}_n\cdot{\bf m}_{n'}.
\end{eqnarray}
Thus weak exchange between the  electrons and the core spins generates 
effective  long-range and sign oscillating effective exchange interaction 
between the core
spins.  

In the opposite case
$J\to \infty$, we can reduce the DE Hamiltonian (\ref{generic}) 
to effective (classic) $t-J$ model 
(see review \cite{izyumov} and references
therein).  To do that, first diagonalize  
the exchange part of the
DE Hamiltonian by choosing
local spin quantization axis on each site in the direction of ${\bf m}$, 
the latter being determined by polar angle $\theta$ and
azimuthal angle $\phi$.  
In this
representation the Hamiltonian  is 
\begin{eqnarray}
\label{hamlocquant}  
H_{nn'} =  -J\sigma^z\delta_{nn'}
+t_{nn'}\left(\begin{array}{cc}a_{nn'} & b_{nn'}\\
b_{n'n}^* & a_{n'n}\end{array}\right),
\end{eqnarray}
where
\begin{eqnarray}
a_{nn'} =  \cos\frac{\theta_n}{2}\cos\frac{\theta_{n'}}{2}
+\sin\frac{\theta_n}{2}\sin\frac{\theta_{n'}}{2}e^{i(\phi_{n'}-\phi_{n})}\nonumber\\
b_{nn'}=\sin\frac{\theta_n}{2}\cos\frac{\theta_{n'}}{2}e^{-i\phi_{n}}-
\cos\frac{\theta_{n}}{2} \sin\frac{\theta_{n'}}{2}e^{-i\phi_{n'}}.
\end{eqnarray}
The next step
is to apply a canonical transformation 
\begin{eqnarray}
H\rightarrow \tilde{H}=e^SHe^{-S}=H+[S,H]+\frac{1}{2}[S[S,H]]+\dots
\end{eqnarray}
which excludes 
all band-to-band transitions. This can be achieved if we chose the operator $S$
in the form
\begin{eqnarray}
S_{nn'}=-\frac{t_{nn'}}{2J}\left(\begin{array}{cc}a_{nn'} & b_{nn'}\\
-b_{n'n}^* & a_{n'n}\end{array}\right).
\end{eqnarray}
We have
\begin{eqnarray}
[S,H^{ex}]_{nn'}=-t_{nn'}\left(\begin{array}{cc}0 & b_{nn'}\\
b_{n'n}^* & 0\end{array}\right)\nonumber\\
\left[S,[S,H^{ex}]\right]_{nn}
=2\sum_{n''}I_{nn''}\left(\begin{array}{cc}
 -|b_{nn''}|^2 & 0\\
0 & |b_{nn''}|^2\end{array}\right),
\end{eqnarray}
where $I_{nn''}=|t_{nn''}|^2/(2J)$.
Keeping terms up to the second order with respect to $t$ (and only site-diagonal
part of the second order terms) we obtain 
\begin{eqnarray}
\label{hamlo}  
\tilde{H}_{nn'} =  -J\sigma^z\delta_{nn'}
+t_{nn'}\left(\begin{array}{cc}a_{nn'} &0\\
0 & a_{n'n}\end{array}\right)\nonumber\\
+\sum_{n''}I_{nn''}({\bf m}_n\cdot{\bf m}_{n''}-1)\sigma^z\delta_{nn'},
\end{eqnarray}
In the second quantization the Hamiltonian obtained  has the form
(ignoring the constant term)
\begin{eqnarray}
\label{ham}
\tilde{H}=\sum_{nn'}t_{nn'}
\left[ \cos\frac{\theta_n}{2}\cos\frac{\theta_{n'}}{2}\right.\nonumber\\
\left.+\sin\frac{\theta_n}{2}\sin\frac{\theta_{n'}}{2}e^{i(\phi_{n'}-\phi_{n})}\right]
d_n^{\dagger}d_{n'}\nonumber\\
+\sum_{nn'}I_{nn'}{\bf
m}_n\cdot{\bf m}_{n'}d_n^{\dagger}d_{n},
\end{eqnarray}
where  $d^{\dagger}(d)$ is the operator of creation (annihilation) 
of a spinless electron.
Thus strong exchange between the  electrons and the core spins generates 
effective 
short-range antiferromagnetic
exchange between the core spins. 

In conclusion, we  formulated the Dynamical Mean Field Approximation
equations
for the double-exchange model with classical spins
for arbitrary electron dispersion law, concentration and 
relation between the  exchange and
the electron bandwidth. In the vicinity of the 
paramagnetic-ferromagnetic transition critical
temperature $T_c$, these equations were reduced to a MF equation, describing a
single spin in an effective field, proportional to the macroscopic
magnetization, and we obtained explicit formula for the $T_c$. 
For the semi-circular electron density of states we explicitly calculated 
the transition temperature as a function of the exchange interaction and
electron concentration. 

This research was supported by the Israeli Science Foundation administered
by the Israel Academy of Sciences and Humanities.

\end{multicols}
\end{document}